# A new physical mechanism for the onset of atomic ionization in an optical field with jolts of phase

M.N. Shneider[a*] and V.V. Semak[b]

a) Department of Mechanical and Aerospace Engineering, Princeton University, New Jersey 08544, USA
b) Virtual Laser Application Design, LLC
* Corresponding author: m.n.shneider@gmail.com

*"No theory ever agrees with all the facts in its domain, yet it is not always the theory that is to blame. Facts are constituted by older ideologies, and a clash between facts and theories may be proof of progress. It is also a first step in our attempt to find the principles implicit in familiar observational notions."*
  Paul K. Feyerabend, Against Method: Outline of Anarchistic Theory of Knowledge (1975, 1993), 39, Chapter 5

**Abstract**
A theoretical model that describes a new mechanism of atomic and molecular ionization in a low intensity electro-magnetic wave (light or laser beam) with the energy of quanta that is lower than required for a single photon ionization is presented. The essence of the proposed physical mechanism is the step-like gain of energy of a bound electron that occurs every time the phase of the electro-magnetic field jolts. Providing there is sufficiently large number of the phase jolts, the summation of the step increases of the electron oscillation energy can render the total energy of the bound electron such that it exceeds the ionization potential.

The physical mechanisms of atomic and molecular ionization by light (we use the term "light" as equivalent to the term "electro-magnetic radiation") are extensively studied. The created theoretical models, that appear to align with experimental data, are considered as established parts of physical optics, plasma physics, and laser physics. These theoretical models can be divided into three types.

The models of the first type describe atomic and molecular ionization in terms of absorption of a single photon with energy exceeding the ionization potential of an atom or a molecule. The absorbed photon transfers its energy and momentum to a bound electron, knocking it out of the atom. Einstein's theory of photoionization is representative of such models. The ionization of isolated atoms or molecules by hard UV radiation, X rays, and γ radiation is described by the models of the first type.

The models of the second type describe atomic and molecular ionization in terms of simultaneous absorption of multiple photons with sum energy exceeding the ionization potential. The multi-photon ionization mechanism assumes that a bound electron acquires energy and momentum via simultaneous absorption of n-photons, and as a result, it is knocked out of the atom or molecule.

Note, that both mentioned models of atomic or molecular ionization assume that the light is a flux of particulates (the photons) and that the absorption of the photons during interaction with an electron bound in an atom results in the ionization. Thus, the models of these two types relate to the corpuscular nature of light dictated by the postulated quantum-mechanical principle of the wave-particle duality [see, for example, a recent review paper [1]).

The third type of the atomic ionization by light relates to the wave nature of light. Such models describe the process of atomic ionization as impact ionization of neutral atoms by free electrons that acquired energy from a constantly irradiating electro-magnetic wave. [2]. The energy transfer from the electro-magnetic wave to the free electron occurs as incremental energy gain during multiple nonionizing collisions with heavy neutrals and ions. After some number of such collisions while irradiated by the electromagnetic wave, the electron can gain kinetic energy exceeding the ionization potential [2, 3]. The following collision between such an energetic free electron and a neutral atom knocks the bound electron out, producing two free electrons. These two electrons will proceed acquiring the kinetic energy from the wave during collisions and, after gaining energy that exceeds the ionization potential, each will ionize other neutral atoms. This third type of ionization by light is called avalanche ionization. Obviously, the light driven avalanche ionization requires a seed of free electrons. The number of the seed electrons required to support avalanche ionization by light is determined by the correlation between the ionization rate and the free electron disappearance rate. The latter is determined by multiple factors such as the recombination with the ions, attachment to the neutrals, diffusion out of the focused light, etc.

The avalanche ionization of gas by intense light (laser "spark") was one of the first effects observed in the mid-1960s when physicists focused a beam of a Q-switched laser in the air. Since then, the detailed theoretical understanding of the mechanism of avalanche ionization of gases by light, or so called "optical breakdown", was developed. The interest was driven by the high practical value of this effect that became the foundation of multiple spectral and acoustic technologies. The ensuing theoretical models addressed an extensive list of the parameters affecting the light-induced ionization. However, one parameter received little to no attention of the researchers – the physical mechanism of production of the seed of free electrons. The co-creator of the avalanche laser-induced breakdown theory, Yurii Raizer, showed in his monograph [2] that the seed electron number density has negligible effect on the rate of ionization, or formation of optical breakdown plasma. This valid result was seemingly misinterpreted by the laser physicists as insignificance of the physics of production of the seed electrons. The trivial and qualitative-only explanations were commonly accepted as the valid mechanisms – cosmic rays, background radiation, and low ionization potential impurities. Lastly, the attitude was universally adopted: "we know the optical breakdown exists, so the seed electrons do exist and who cares how they are produced". Indeed, the rate of formation of the optical breakdown plasma is a weak function of the seed electron number density. However, whether a laser can produce optical breakdown or not is a binary function of the seed electron existence, i.e., no seed electrons – no breakdown due to the avalanche ionization. Then, the seed electron mystery requires rigorous study and deserves a comprehensive theoretical model.

Let's demonstrate the above point. Easy, estimated calculations show that none of the "obvious" and "trivial" mechanisms from the commonly repeated list is capable of producing background electron density that corresponds to at least one seed electron within the focal volume of a laser beam. Indeed, if the background electron number density is $10^2$ cm$^{-3}$, then for the focal volume of a nanosecond laser ~$10^{-5}$ cm$^3$ (20 μm radius and 10 mm Rayleigh length) the breakdown probability is ~$10^{-3}$ or 0.1%. In other words, for a laser operating at 1 kHz pule repetition rate, the breakdown will be observed once per one second, independent of other laser parameters. To give another practical perspective, for a single pulse laser it will take 1000 pulses to observe optical breakdown one-time. A high pulse energy laser used in laser induced breakdown spectroscopy (LIBS) can typically generate one pulse per minute. Then, the scientists who use LIBS for material characterization, would record one spectrum per two days, assuming an 8-hour workday. Anyone who has experience with lasers knows that once the laser parameters exceed the threshold, the probability to produce breakdown in one laser pulse is practically 100%. Such perspective moves the model of the physical mechanism of the seed electron production from the "trivial and unimportant" bin to the "highly important" bin. Let us now present some quantitative considerations that support our assessment of importance of understanding the physics of production of seed electrons for light-induced avalanche ionization of a gas.

The simple estimates using well-known equations [4] show that seed electron generation by multiphoton or tunneling ionization of atoms requires light intensities that exceed ~$10^{18}$ W/m². Such beam intensity can be achieved in the focused beams of the femto- and pico-second pulse lasers. However, in case of a nanosecond pulse laser, the optical breakdown is known to occur at light intensities much lower than $10^{17}$ W/m². At such low light intensities, the probability to produce seed electrons as a result of multiphoton ionization/tunneling is practically zero.

Similarly, the background radioactive ionizing radiation cannot be a source of the initial electrons required for development of avalanche ionization breakdown. The sources of natural background radiation are cosmic radiation, solar radiation, external terrestrial sources, and radon. For example, in noble gases at 300K and arbitrary pressure, the ionization rate due to background radiation is $G_i \approx 6 * \frac{10^6 p_{[Torr]}}{760}$, m⁻³s⁻¹ [5]. The loss of electrons produced by the background radiation is determined mainly by the process of binary dissociative recombination of electrons and dimer molecular ions, formed in three-body collisions of atomic ions with neutrals, with the rate $G_r = \beta_d n_e^2$, where $n_e$, m⁻³, is the number density of free electrons and the recombination constant $\beta_d \approx 10^{-12} \left(\frac{300}{T_{e,[K]}}\right)^{0.61}$, m⁻³s⁻¹ [6]. For normal atmospheric condition, the estimate of electron density corresponding to the threshold, when the ionization rate equals the recombination rate, gives $n_e = \sqrt{\frac{G_i}{\beta_d}} \sim 2.4 * 10^9$, m⁻³. In the case of a laser beam focused into a spot with σ = 20 microns radius with the caustic length of $L_R$ = 1 cm, the number of initial electrons, $N_e = n_e \pi \sigma^2 L_R$, is approximately 3*10⁻². Thus, just one seed electron produced by the background radiation is found in the volume that exceeds the typical laser beam focal volume by at least 30 times. Then, nanosecond laser optical breakdown must always be a stochastic process with the occurrence probability that is no better than ~ 1/30. Practice shows that this is not the case.

Finally, the presence of easily ionizable, low concentrations of organic impurities can significantly reduce the optical breakdown threshold. It is possible to estimate the probability of finding at least one atom of a species with sufficiently low ionization potential in the focal volume of a laser beam focused in air. However, experiments show that energetic and long duration laser pulses produce optical breakdown in high purity gases with 100% probability. Thus, the mystery of the source of the seed electrons for these conditions cannot be resolved by the presence of impurities.

Here, we address the mystery of the seed electrons by presenting our theoretical model of atomic ionization by light with jolt-jittering phase. This new theory indicates that in addition to the three previously mentioned mechanisms of atomic/molecular ionization by light, there is another universal mechanism. This new mechanism explains the generation of the initial electrons required for avalanche ionization of gas in focused light beams. The proposed mechanism can be understood in terms of the energy gain by a system that experiences transition from regular to chaotic motion under the effect of an external force that undergoes jolt-like changes of its value or direction [7]. An illustrative example of such a process is the Fermi-Ulam mechanism of gain of energy of the charged particles during reflection from magnetic mirrors [7]. This mechanism is evoked to explain stochastic heating of electrons in RF capacitive discharge [8] and instability of the planetary orbits [9].

Let us describe our new theoretical paradigm for gain of energy of an electron bound in an atom and exposed to low intensity light. For a typical laser, the angular frequency of laser radiation $\omega_L$ is much lower than the frequency of natural oscillations of a bound electron, $\omega_0$. If the phase of the optical field changes randomly, then the bound electron incrementally acquires the oscillation energy during the phase jolts. Every laser experiences fluctuation of the phase of emitted radiation. The characteristic time between the phase joilts determines the linewidth of spectral emission and the coherence length (see, for example, [10]) $L_{coh} = c\tau_{coh} = c/\pi\Delta\nu$, where $\tau_{coh}$ is the time over which a propagating wave may be considered coherent, meaning that its phase is, on average, predictable. Note that $\Delta\nu$ is the full width at half-maximum linewidth (optical bandwidth).

For the cw lasers operating in a single mode regime, the coherence time and the coherence length are comparatively long, and the phase jolts are relatively less frequent. When such a highly coherent laser beam interacts with an atom, the gain of the electron oscillation energy will be compensated by losses due to atomic or molecular collisions. Then, the atomic ionization doesn't take place and the seed electrons cannot be produced. For such lasers with highly monochromatic radiation output, the optical breakdown should occur at higher radiation intensities when multiphoton ionization becomes possible.

Typical pulsed nanosecond and picosecond lasers have relatively broad bandwidth, and thus the coherence length is relatively small (from a few millimeters to centimeters [11]) and the coherence time is $\tau_{coh} \ll \tau_{pulse}$. Then, during the laser pulse, the phase of the radiation jolts hundreds or thousands of times. Each phase jolt results in energy transfer from the electro-magnetic wave increasing the energy of the bound orbital electron. For a higher laser intensity, i.e. higher optical electric field amplitude, the incremental amount of energy acquired by the electron in a single phase jolt is larger. At a certain threshold intensity, the sum of the increases of the electron oscillation energy corresponding to each phase jolt becomes sufficient for the electron to overcome the bonding potential and become free.

It easy to see [12] that the average kinetic energy of oscillations of a bound electron acquired during one phase jolt is much smaller than the energy of a photon: $\frac{m\dot{r}^2}{2} \ll \hbar\omega_L$. It is reasonable to postulate that [12] scattering of a photon on an individual atom takes place only when the energy of induced oscillation of the atomic bound electron is no lesser than the energy of the photon. Therefore, in the laser beam interaction with a single atom, the atom cannot produce photon re-emission, i.e. scattering, if the laser wavelength is in the visible – IR part of the spectrum [12]. Thus, in the intervals between the phase jolts, the oscillation energy acquired by the electron is not re-radiated and it remains stored in the electron's natural oscillations.

As an example, let us consider the semi-classical approximation of interaction of a hydrogen atom with an electromagnetic wave. For this purpose, we apply our recently published Atomic Oscillator Model [13-15]. In this model the electron displacement from the stationary Bohr's orbit of the s-state of a hydrogen atom is provided by the solution of the following equation of motion:

$$\ddot{r} + \frac{2U_0 r_0^2}{m_e}\left(\frac{1}{r_0 r^2} - \frac{1}{r^3}\right) - \frac{\xi}{m_e}\dddot{r} = -\frac{e}{m_e}E(t), \tag{1}$$

where $\xi = e^2/6\pi\varepsilon_0 c^3$, $\varepsilon_0$ is the vacuum permittivity, $c$ is the speed of light, $m_e$ and $e$ are the electron mass and charge, $r_0$ is the radius of the equilibrium (stable) orbit of electron, and $U_0$ is the depth of effective electron potential.

Using equation (1), one can compute the increase of the oscillation energy of a bound electron. This increase of electron energy can be computed as the work produced by the field on the electron, i.e., as the time integral of the product of the electron charge, electron velocity, and forcing electric field. If the field is a sinusoidal function and the time interval of integration is taken within the established electron motion, then it is usually said that the work of the field on the electron is zero. Thus, in a continuous electro-magnetic wave, an electron gains no energy. This is widely known. However, what does not appear to be widely known is that an electro-magnetic wave with varying amplitude or phase produces nonzero work on an electron, i.e., it causes increase of the electron energy. Under most familiar light-atom interaction conditions (when the photon energy is small, i.e., when the wavelength is below the UV part of spectrum) this increase of electron energy is much smaller than the energy of a single photon. For example, in the case when an electro-magnetic wave is turned on instantaneously and kept at constant amplitude, the electron energy gain is approximately equal to the maximum kinetic energy during one full oscillation. For the commonly familiar conditions of laser irradiation, the energy of induced oscillations of the electron is less than $10^{-5}$ - $10^{-3}$ of the photon energy, depending on laser irradiance. It appears to be negligibly small, and possibly, this is the reason why such consideration is never taught in the physics courses.

In cases where the field amplitude is changing, or the field phase experiences jolts, the electron will continue gaining its oscillation energy after the wave is "turned on". The physical mechanism of this gain is straightforward, and the mathematical model is simple. In spite of the clarity of the physical model and obvious importance of the bound electron energy gain in the field with varying amplitude or phase jolts, such a mechanism was previously never researched, neither theoretically nor experimentally.

Let us initiate such research here by conducting analysis of the limiting case when the laser is turned on rapidly during time, $t_p$, much shorter than the laser pulse duration. The laser intensity remains constant after the initial "turning on" phase, and the third term in the left-hand side of equation (1) is assumed to be zero (i.e. the re-radiation due to the forced electron motion is assumed as zero). The reason for the latter assumption is to keep our physical model consistent with the postulates of quantum mechanics [12]. In contradiction to all current physical optics textbooks, we propose that the re-emission of the electro-magnetic wave cannot take place until the energy of forced electron oscillations exceeds the energy of the re-emitted photon. Hence, we zero the "re-emission," or "scattering," reflected in the third term on the left-hand side of equation (1). The detail considerations behind this can be found in our recent work [12].

The exact analytical solution of equation (1) is impossible without simplification. We will avoid the approximation (linearization) of equation (1) typically found in the textbooks and other publications. Indeed, we look for a high accuracy solution since the effect of energy gain is relatively small. Thus, the numerical solution of equation (1) is our only option.

Let us represent the electric field in the right side of equation (1) as a harmonic function with varying amplitude and randomly jolted phase:

$$E(t) = E_a(t)\cos(\omega_L t + \Delta\varphi), \tag{2}$$

where $E_a(t)$ is the laser radiation field amplitude, $\Delta\varphi(t) = 2\pi\xi(t)$ is the random phase shift, where $0 < \xi(t) < 1$ is the random number that jumps after $N$ laser periods, $T_L = 2\pi/\omega_L$. For the sake of an example simulation let us express the ampltude, $E_a$, with the following function:

$$E_a(t) = E_{a,0}\left(1 - exp\left(-\frac{t}{t_p}\right)\right), \tag{3}$$

where the characteristic rise time of the "pulse" is $t_p$. In the simulation we will assume the value of this rise time as $t_p$ = 30 fs.

Solving equation (1) numerically provides the radial position of a bound electron, $r(t)$. Computing the time derivative of the electron position, i.e., computing its velocity, provides the kinetic energy of the electron oscillations induced by the field of the incident electro-magnetic wave

$$\varepsilon_{osc}(t) = \frac{m_e[\dot{r}(t)]^2}{2}. \tag{4}$$

This oscillation energy that the electron acquires from the light wave varies in time, undulating within the light period, $T_L = 2\pi/\omega_L$, from a maximum value down to zero. The average kinetic energy of the electron during one oscillation period of the field reamins constant in time if the phase and amplitude of the field remains constant in time. The reader can find additional information derived for a particular expression of the effective potential of an electron that we used in our calulations [12-15]. In the case when the phase of the field oscillation is constant and the amplitude is increasing from zero to some value that remains constant (for example, as described by equation (3)), the oscillation energy maximum and its average value during the field period, increase from zero to some constant value. Note, that the largest value of the established amplitude of oscillation energy is achieved for a free electron. In the case of a bound electron, the steady state average oscillation energy is noticeably smaller and decreases with the increase of difference between the forcing frequency, $\omega_L$ in our case, and the electron's natural oscillation frequency, $\omega_0$. All this is well known and described in theoretical mechanics testbooks, for example, [16].

For simplicity, but without loss of generality, we consider the interaction of a hydrogen atom with the radiation of the 2nd harmonic of a pulsed Nd:YAG laser with a wavelength of λ=532 nm ($h\nu$ = 2.33 eV). As in our previous works we model electron motion assuming the Bohr – Sommerfeld model of an atom. According to the Bohr – Sommerfeld model, an electron in a hydrogen atom, in a spherically symmetric s-state and at the lowest energy level, moves in the effective potential with depth $U_0 = I_i$, where $I_i$=13.6 eV is the ionization potential from the ground state. Since there is no reliable and unambiguous information about the values of atomic radii, let us assume here the value $r_0 = r_B$, where $r_B$ = 52.9 pm is the Bohr radius. Note, the allowable region of the possible radius of a hydrogen atom could range from the Bohr's radius to the Van der Waals' radius ~ 120 pm. As far as we know, no one has made direct measurements of the effective radius of a hydrogen atom. As we will see below, the value of the radius of a hydrogen atom can be estimated from the observed spectra of dipole radiation generated by a short pulse of electromagnetic radiation. Thus, the latter could be an excellent tool for the direct characterization of atoms and molecules.

Let us present the results of the numerical solution of equation (1) for forced electron deviations from the steady orbit and electron velocity. Using the latter result and multiplying it by the electric charge and the value of the electric field gives us the instantaneous power that the wave exerts on the electron. Finally, running the time integration of the instantaneous power exerted on the electron from the beginning of interaction to the current time value gives the current value of the electron oscillation energy. Note that according to conventional, educational physics, the value of the instantaneous power varies from positive, when the electron gains oscillation energy from the electro-magnetic wave, to mirror negative, when the electron "returns" the oscillation energy back to the electro-magnetic wave. The majority of the physics community believes that the net value of the energy of electron oscillation forced by an electro-magnetic wave is zero. It is uncommon knowledge that, in the case of variable amplitude or phase of electric field, the energy of electron oscillations is non-zero and depends on the "history" of the field. A detailed description of energy transfer to a medium from a signal, discussed on a macroscopic level, can be fould in the extraordinary works of Brilloun and Sommerfeld [17,18]. These theoretical works are practically unknown; indeed, in 2003 one of the authors borrowed these books in a library and found that they were previously checked out only once in 1956 and 1963, respectively. The authors are aware of only one microscopic consideration of an electron acquiring a non-zero amount of oscillation energy from a pulse of an electro-magnetic wave [3]. In this book, an estimate is given only for a free electron – it is estimated that the oscillation energy acquired from a laser beam approximately equals the average energy of one oscillation [3]. We further develop and extend the theoretical model to include the case of a bound electron on the atomic/molecular level, assuming that the forced electron oscillation occurs in an effective potential [12-15] and the electric field experiences phase jolts where its amplitude is rapidly changing. Below we provide the results of the computed velocity and energy of the electron oscillation, forced by the electric field with rising amplitude and randomly jolted phase, described by equations (2) and (3).

For a maximum laser beam intensity $I_L = 3.3 \cdot 10^{16}$ W/m$^2$, corresponding to the maximum amplitude of the electric field $E_a = 5 \cdot 10^9$ V/m, and under the condition when the phase of the electric field remains zero, the computed velocity of forced motion of the electron and corresponding oscillation energy are given in Figure 1 a) and b), correspondingly. For reference, note that for the laser pulse energy of 100 mJ, with radius of the focused beam of 10mkm radius at the 1/e level of maximum intensity and laser pulse duration of 10ns, the maximum laser beam intensity is $3.18*10^{16}$ W/m$^2$. Under these conditions the bound electron exhibits well-known behavior. Forced by the laser pulse with rise time of 30 fs, the electron undulates around a steady orbit with amplitude and velocity (Figure 1 a) that increase during the transient stage, reaching a steady value after approimately 200 fs. The corresponding computed electron oscillation energy increases during the same time from zero to a steady value of approximately $9.5*10^{-6}$ eV, which is approximately the energy of one electron oscillation.

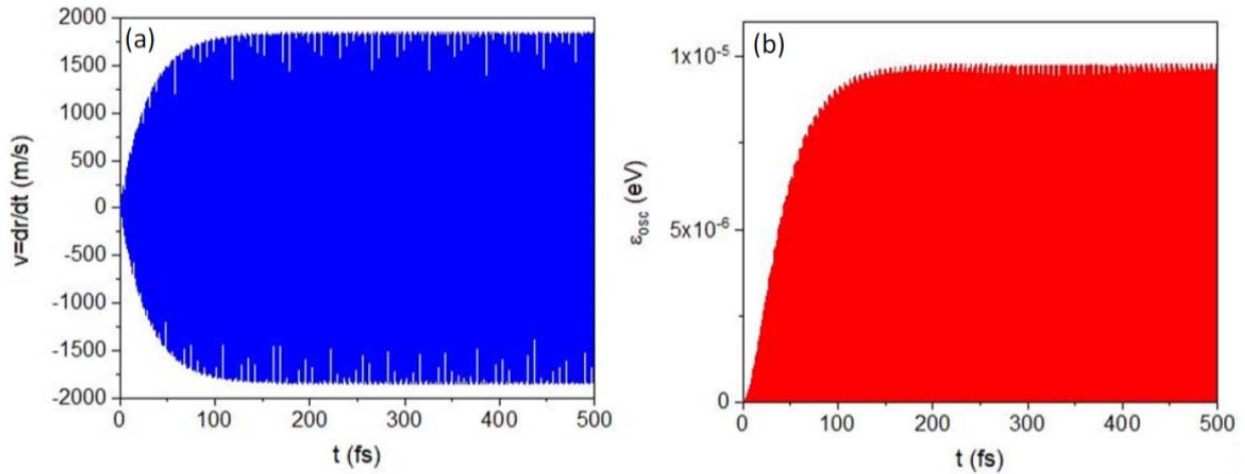

**Fig. 1.** (a) - Oscillatory electron velocity and dependence of the kinetic energy of oscillations on time for a laser with radiation in which there is no phase failure (radiation with an infinite coherence length). (b) - Electric field amplitude $E_a = 5 \cdot 10^9$ V/m, corresponding intensity, $I_L = 3.3 \cdot 10^{16}$ W/m$^2$.

For the same maximum laser intensity and same temporal shape of the front of the laser pulse, and assuming that the field phase experiences random jolt every 50 periods of laser oscillation, the computed electron velocity and oscillation energy are shown in Figures 2 a) and b), correspondingly. As one can see, if the phase of the laser field exhibits random jolts, the maxima of the electron oscillation energy experience substantial and rather chaotic fluctuations, unlike in the case of constant phase. The maxima of the oscillation energy fluctuate in the range between 0.1eV and 0.3 eV. Thus, the "average" oscillation energy of the bound electron, subjected to the field with jolted phase, is at least $10^4$ times larger than if the field phase were constant. We observed such dynamics of fluctuations of the maximum oscillation energy remaining below 0.3 eV for at least 120 ps of the computation time. This leads us to project that for this lower light intensity, the maximum energy of oscillation will remain lower than 0.3 eV for even longer durations.

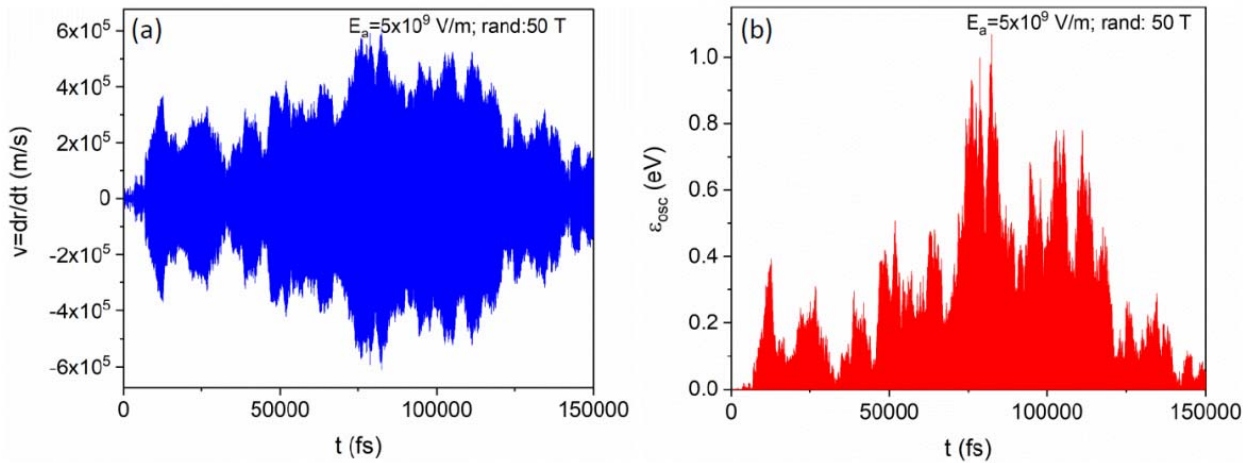

**Fig. 2.** The same as shown in Fig. 1, but in the calculation, it is assumed that every 50 periods there is a random jolt of the phase of the laser radiation.

The next set of computed results demonstrates what happens when the light intensity is increased (for example, see the 4 times higher light intensity shown in Figures 3 and 4). If the phase jolts are not present and the phase of the field is constant, after the transient stage the maximum electron oscillation energy remains steady at the value of approximately the energy of a single oscillation, as shown in Figure 3. Indeed, compare oscillation energies of $0.95*10^{-5}$ eV corresponding to the light intensity of $3.3 \cdot 10^{16}$ W/m$^2$ in Figure 1 and $3.8*10^{-5}$ eV corresponding to four times higher light intensity of $1.33 \cdot 10^{17}$ W/m$^2$ in Figure 3. However, when the electric field phase jolts, the maximum of the oscillation energy of the electron is increasing, see Figure 4. At first, the increase is relatively slow, from zero to ~ 4 eV; however, at later stages of calculation an explosive increase of the electron oscillation energy is observed, up to the value corresponding to the electron escape from the potential well, i.e., atomic ionization by light takes place.

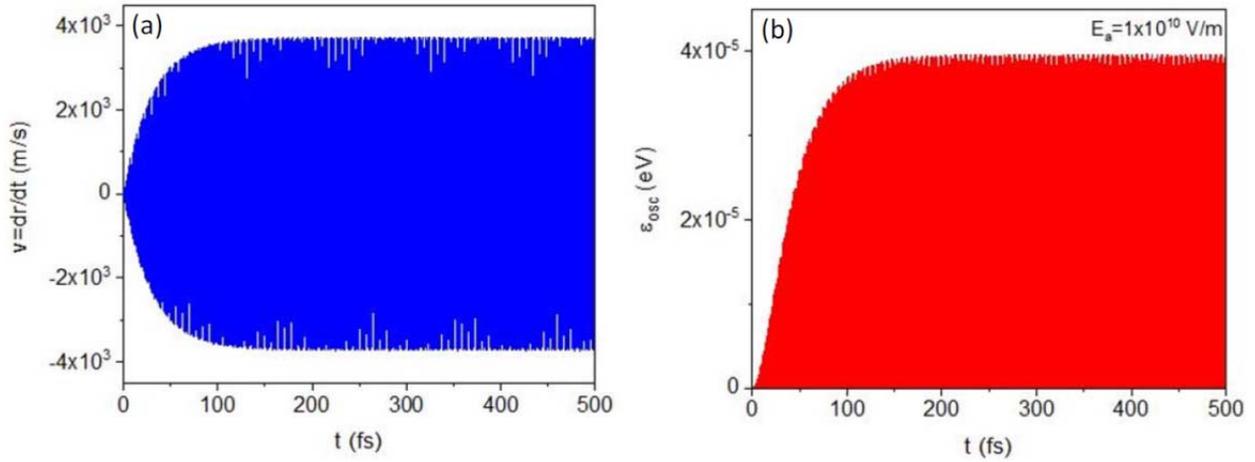

**Fig. 3.** The same as in fig. 1, but for the electric field amplitude $E_a = 10^{10}$ V/m, the corresponding intensity, $I_L = 1.33 \cdot 10^{17}$ W/m²

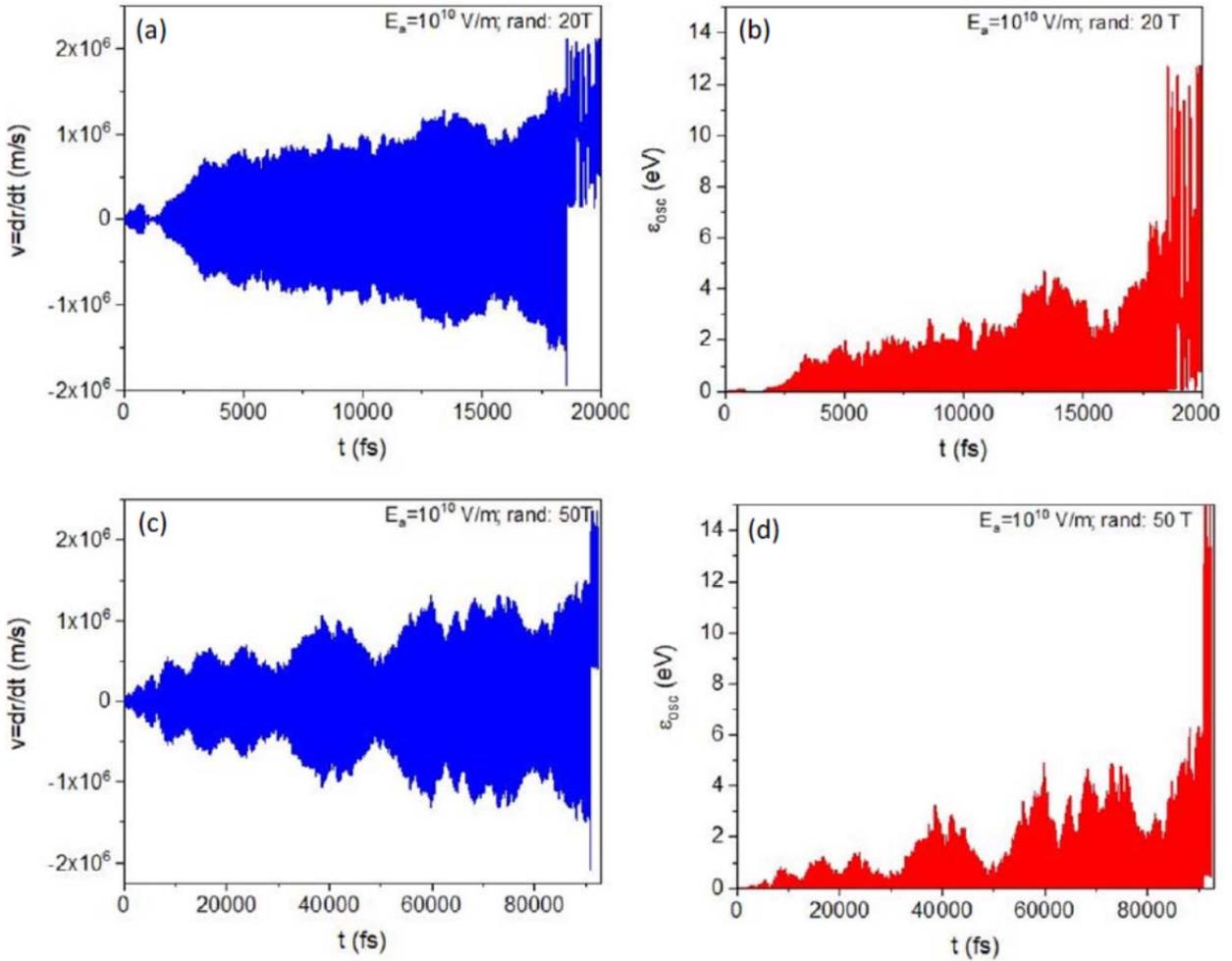

**Fig. 4.** The same as shown in Fig.3. (a), (b) calculations assume that every 20 periods there is a random failure of the phase of the laser radiation; (c), (d) phase failure occurs every 50 cycles.

The time that elapses from the beginning of the light pulse to the moment of explosive increase of the maximum oscillation energy of the bound atomic electron depends on how frequently the phase jolts take place. For example, if the time between the phase jolts is 50 periods of the electro-magnetic wave, the rapid increase of electron oscillation energy, i.e., ionization, starts at 90 ps after the beginning of the laser pulse (Figure 4 b,c), and if the time between the jolts is 20 periods the ionization time decreases to 1.7 ps (Figure 4 a,b).

The simulation results presented in Figure 5 provide an enhanced view of the dynamics of the phase jolts (Figure 5a) and corresponding step increase of the amplitudes of the oscillation velocity and the energy, shown in Figures 5b and c, correspondingly. One can see at the moment of phase jolt indicated by the arrow in Figure 5 a), the amplitudes of both velocity and energy of the oscillations of the bound electron experience step-like increase, See Figure 5 b) and c), correspondingly. For additional clarification, note that the oscillation velocity and energy undulate at the electron's natural frequency (higher frequency), and the amplitude of these undulations is modulated at the lower driving force frequency, i.e., the laser radiation frequency (Figure 5 b and c).

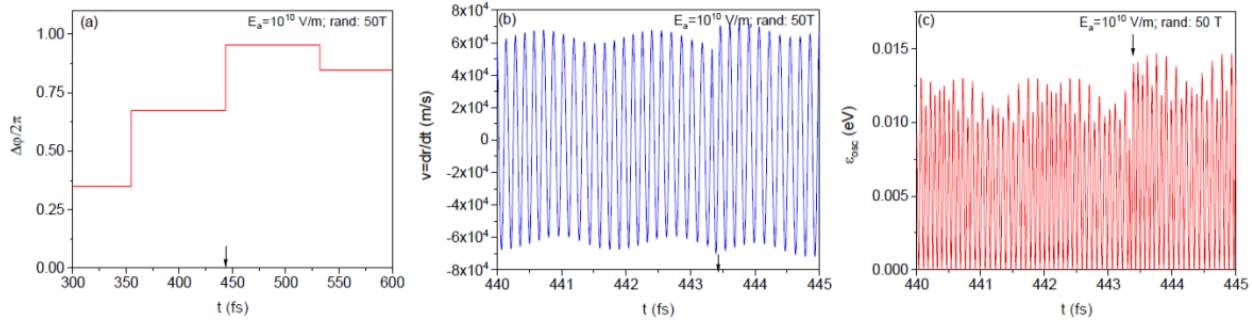

**Fig. 5.** (a) Random phase change occurring every 50 periods (corresponding to the solution shown in Fig. 4c,d); (b), (c) an example of the oscillatory velocity and kinetic energy in the vicinity of the moment of phase change indicated by the arrow in Fig. (a).

In conclusion, let us present some information on the currently available laser systems in order to provide a broader context of our theoretical work and numerical simulations. First, the theoretical model we used for electron dynamics is our recently proposed Atomic Oscillator Model [13] which is drastically different from the commonly used Lorentz Oscillator Model [19,20]. Unlike Lotentz' model, our model is founded on paradigms that are consistent with modern quantum mechanics. In particular, in our model the atomic electron resides in a realistic potential that is utterly different from the quadratic potential assumed in Lorentz' model. Then, a reader might assume that the result of the numerical simulations is due to the noticeable nonlinearity of the "returning" force acting on the electron. It is easy to see theoretically that an electron forced to oscillate by an electro-magnetic wave will incrementally gain energy every time the phase of the field is jolted for any shape of the potential, including Lorentz' quadratic potential. Furthermore, our theory shows that a free electron will gain kinetic energy in a similar manner if the phase of the wave experiences jolts. Second, let us present relevant specifications of the lasers used to produce optical breakdown in gases. Of particular interest for our discussion is the laser linewidth, i.e., the coherence time. For example, the lasers used in LIBS and manufactured by Amplitude Laser, Inc have typical linewidth of 1 cm$^{-1}$ or 30 GHz (see spects for Inlight or Powerlight-DLS [12]). The coherence time that corresponds to the specified laser linewidth is 5.31 ps. Then, during 10 ns laser pulse the phase jolts occur 1885 times. In our computations, in order to obtain the result in some reasonable time (~20 hours per case), we used the coherence time of 50 laser oscillation periods of a wavelength with 532 nm, i.e., the coherence time was 88. 33 fs. Thus, the linewidth of our virtual laser was 1800 GHz, 60 times larger than the linewidth of the high pulse energy nanosecond lasers offered by Amplitude Laser, Inc. Note that the optional models of these lasers that utilize an injection seed have an even narrower specified linewidth of 0.003 cm$^{-1}$, or 90 MHz.

Does such a drastic difference between the linewidth used in computations and the linewidth typical for a commercially offered laser have an effect on the time required to produce ionization? Our numerical "experiments" demonstrate that the effect would be significant. For example, comparing the computation results for the same amplitude of the optical electric field $E_a = 10^{10}$ V/m, and the corresponding intensity, $I_L = 1.33 \cdot 10^{17}$ W/m$^2$, it is easy to see that the onset of rapid increase of electron oscillation energy leading to ionization occurs much earlier for the shorter coherence time, i.e. for a broader laser linewidth. Then for the real laser, the ionization time will be later than the one computed, assuming the broader laser linewidth, or a higher laser intensity would be required to achieve ionization within the same time. The comparison of the results of additional numerical experiments with the measured threshold intensities and the breakdown onset times could easily verify our theory.

It is relevant here to mention that the effect of the atomic and molecular collisions on the increase of the average oscillation energy of a bound electron is absolutely the same as the effect of the jolts of the phase of the light field. In the air at atmospheric pressure, the collision rate is ~$10^{12}$ s$^{-1}$. Then, even if the laser linewidth is as small as several MHz, the electron oscillation will be jolted in average once per one picosecond, resulting in the gain of the average oscillation energy, as described here.

Thus, we present in this work a theoretical model that describes the gain of the energy of a bound electron in an atom/molecule, up to overcoming the ionization potential. The increase in the kinetic energy of the electron oscillatory motion occurs due to the jolts of the phase of the radiation fields. The effect of the phase jolts on the motion of a bound electron was never considered prior to our work. However, the realistic model of the interaction of light with atoms and molecules must include the dynamics of the field amplitude and phase, since every practical radiation source has a finite (usually, relatively short) coherence length and coherence time. The proposed and previously unknown physical mechanism of the energy gain of a bound atomic electron has superficial resemblence with the mechanism described by the Fermi- Ulam model of the increase of mechanical energy of a particle that collides elastically between a fixed wall and a moving one, each of infinite mass. This described mechanism of the bound electron energy gain in an electro-magnetic wave is a new addition to the other known mechanisms of material ionization in an optical field, such as the avalanche and multiphoton ionization. Additionally, the proposed mechanism represents a feasible model for the excitation of the electronic oscillations in an atom, or a molecule that produces subsequent emission of radiation. In this case, the ionization does not take place and the excited electron remains bound.

**Acknowledgments**

The authors are extremely grateful to Heather Danner for the assistance in editing the text of the article, her enthusiastic support, and multiple and valuable discussions of the implications of our research in fundamentals of physical optics that lead to this particular project.

**References**
1. L. Gallmann, I. Jordan, H. J. Wörner, et. al. (2017) Photoemission and photoionization time delays and rates, Structural Dynamics 4, 061502; https://doi.org/10.1063/1.4997175
2. Yu.P. Raizer, Laser-Induced discharge phenomena (Consultants Bureau: New York, 1977)
3. Yu. P. Raizer, Gas Discharge Physics (Springer, New York, 1991)
4. See https://en.wikipedia.org/wiki/Ionization
5. S. I. Yakovlenko, (2004) Mechanism for the Streamer Propagation toward the Anode and Cathode due to Background Electron Multiplication, Technical Physics, 49, 1150
6. Y.-J. Shiu, M. A. Biondi, (1978) Dissociative recombination in argon: Dependence of the total rate coefficient and excited-state production on electron temperature, Phys. Rev. A 17, 868
7. A.J. Lichtenberg, M.A. Lieberman, Regular and Chaotic Dynamics, 2nd edition (Springer Science+Business Media, LLC, New York 1992)
8. V. A. Godyak, (1972) The statistical heating of electrons by oscillating boundaries of the plasma, Sov. Phys.-Tech. Phys., 16, 1073; (1998) M.A. Lieberman, V.A. Godyak, From Fermi Acceleration to Collisionless Discharge Heating, IEEE Trans. Plas. Sci., 26, 955
9. K. Batygin, R. A. Mardling, and D. Nesvorný, The Stability Boundary of the Distant Scattered Disk, (2021) The Astrophysical Journal, v. 920, n. 2
10. M. Young, Optics and Lasers: Including Fibers and Optical Waveguides (Advanced Texts in Physics) 5th ed. (Springer, Berlin, New York, 2000)
11. See website of the Amplitude Laser company - https://amplitude-laser.com; Specifications of the lasers Inlight and Powerlight-DLS
12. V.V. Semak and M.N. Shneider, (2021) Different Perspective on Blue Sky Theory: Theory of Single-Photon Scattering on Bound and Free Electrons, arXiv preprint arXiv:2107.06143
13. V.V. Semak, M.N. Shneider, (2017) Invicem Lorentz Oscillator Model, arXiv: 1709.02466
14. V.V. Semak and M.N. Shneider, (2020) Predicted response of an atom to a short burst of electromagnetic radiation, OSA Continuum, 3(2), 186
15. V.V. Semak, M.N. Shneider, (2019) Analysis of harmonic generation by a hydrogen-like atom using quasi-classical non-linear oscillator model with realistic electron potential, OSA Continuum, 2, 2343
16. Douglas Cline, Variational Principles in Classical Mechanics (University of Rochester, 2021; 3rd edition) ISBN: 978-0-9988372-3-9 e-book (Adobe PDF)
17. L. Brilloun, Wave Propagation and Group Velocity (Academic Press, New York and London, 1960) ISBN-10:1483253937
18. L. Brilloun, Science and Information Theory, (Academic Press Inc., New York, 1956) ISBN-10: 0486497550
19. Lorentz H.A., The Theory of Electrons and the Propagation of Light, Nobel Lecture, December 1 1, 1902; https://www.nobelprize.org/prizes/physics/1902/lorentz/lecture/
20. Almog, I. F., Bradley, M. S., Bulovic, V. (2011). The Lorentz Oscillator and its Applications, Department of Electrical Engineering and Computer Science, Massachusetts Institute of Technology; https://ocw.mak.ac.ug/courses/electrical-engineering-and-computer-science/6-007-electromagnetic-energy-from-motors-to-lasers-spring-2011/readings/MIT6_007S11_lorentz.pdf